\documentclass[runningheads]{llncs}

\usepackage{amsmath}
\usepackage{graphicx}
\usepackage{tikz-network}

\begin{document}

\title{Computing Curved Area Labels in Near-Real Time}
%
%
\author{
Filip Krumpe \and
Thomas Mendel}
\authorrunning{F. Krumpe, T. Mendel}
%
\institute{University of Stuttgart, Stuttgart, Germany\\
Institut of Formal Methods in Computer Science - Algorithmic Group\\
\email{\{krumpe,mendel\}@fmi.uni-stuttgart.de}}
\maketitle              

\begin{abstract}
    
    In the Area Labeling Problem one is after placing the label of a geographic area.
    Given the outer boundary of the area and an optional set of holes.
    The goal is to find a label position such that the label spans the area and is conform to its shape.

    The most recent research in this field from Barrault in 2001 proposes an algorithm to compute label placements based on curved support lines.
    His solution has some drawbacks as he is evaluating many very similar solutions.
    Furthermore he needs to restrict the search space due to performance issues and therefore might miss interesting solutions.

    We propose a solution that evaluates the search space more broadly and much more efficient.
    To achieve this we compute a skeleton of the polygon.
    The skeleton is pruned such that edges close to the boundary polygon are removed.
    In the so pruned skeleton we choose a set of candidate paths to be longest distinct subpaths of the graph.
    Based on these candidates the label support lines are computed and the label positions evaluated.

\end{abstract}


\section{Introduction}

    In todays highly mobile world everything is about being able to orient oneself in new environments.
    Not for nothing are navigation apps an integral component of todays smart devices.
    In contrast to former static maps todays maps are no longer restricted to fix map scales.
    Hence they allow to explore virtually an arbitrary amount of information by interactively panning and zooming the map view.
    This means that map can no longer be created by hand but need to be created automatically.
    Especially in view of the fact that maps might be highly customizable in the feature.

    The labelling of places in maps is a vast problem in which one aims at annotating text to maps.
    The main challenges are to determine which elements or places to label and where to place the label in order to help the user to identify the corresponding object.
    In static maps, e.g. paper maps, one can tackle this problem by spending a huge amount of effort, may it be human editing or computational power.
    In the meantime the focus in map usage has shifted from static maps to interactive ones.
    These maps allow to pan, rotate, and zoom a map view continuously.
    It is impossible to deal with the infinite possibilities that arise in this way with static labellings.
    Hence one has to find schemes to automatically determine a proper labelling for a given setting in real-time.

	\begin{figure*}
		\centering
        \includegraphics[trim={0 200 0 200},clip,width=0.98\textwidth]{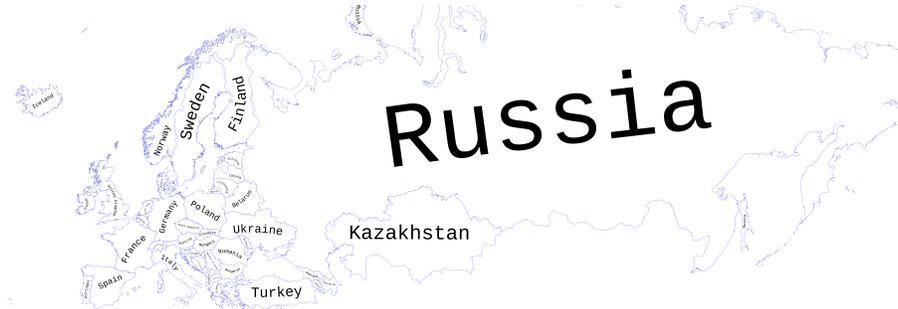}
		
		\caption{Political map of the European countries including Russia with the automated labelling process we propose.}
		\label{fig:europeRuss}
    \end{figure*}

    In map labelling a distinction is made between the labelling of point-like data (i.e. points of interests), one-dimensional data (streets, rivers) or two-dimensional data (lakes, countries).
    But with the focus on interactive display, each of these elements corresponds to a real world place which has a spatial extent.
    At an sufficiently large map scale a place should be labelled within its boundaries.
    This is what we are focussing in the work at hand.

\begin{figure}[ht]
        \centering
        \includegraphics[trim={0 200 0 200},clip,width=0.47\textwidth]{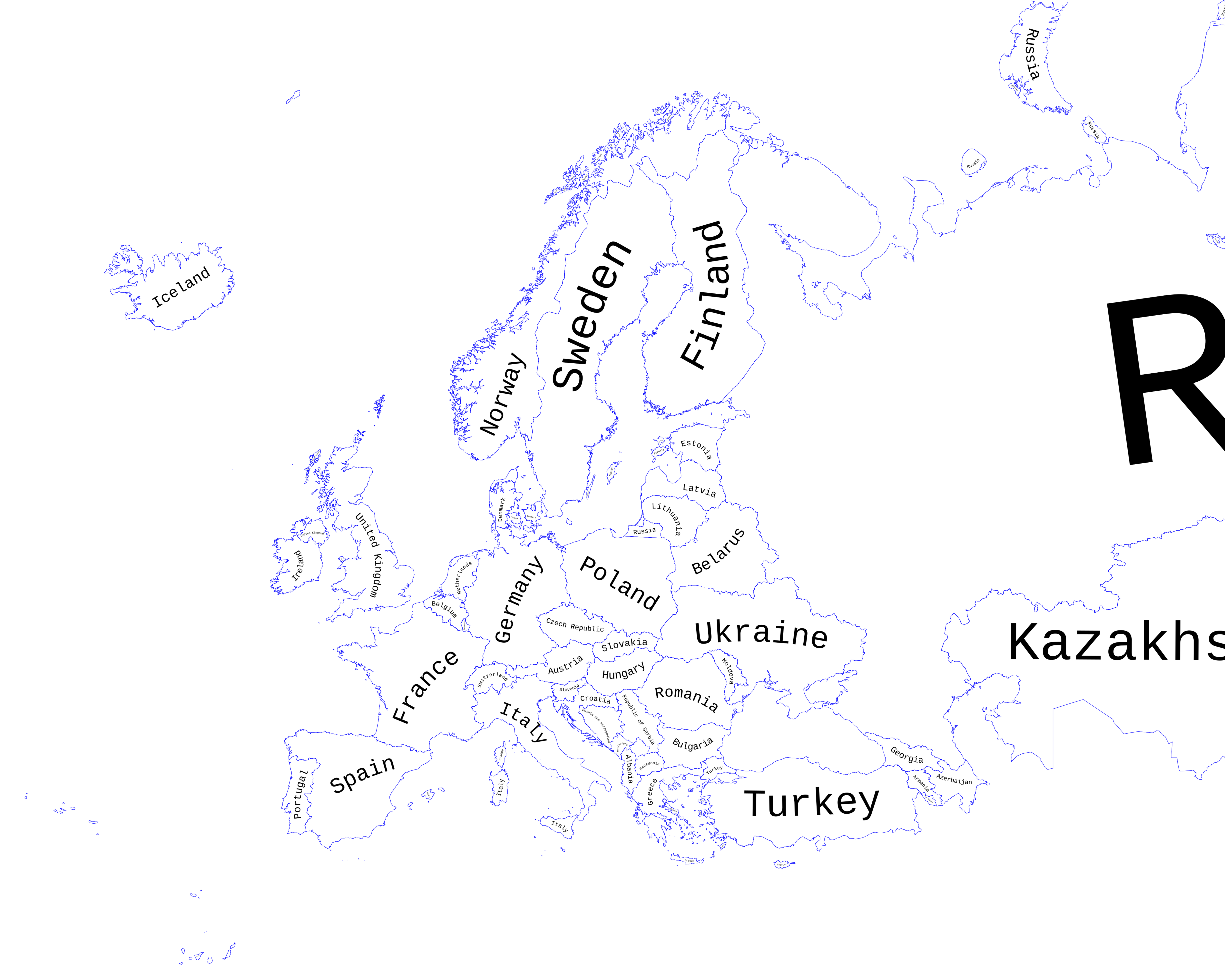}
        \hfill
        \includegraphics[trim={0 200 0 200},clip,width=0.47\textwidth]{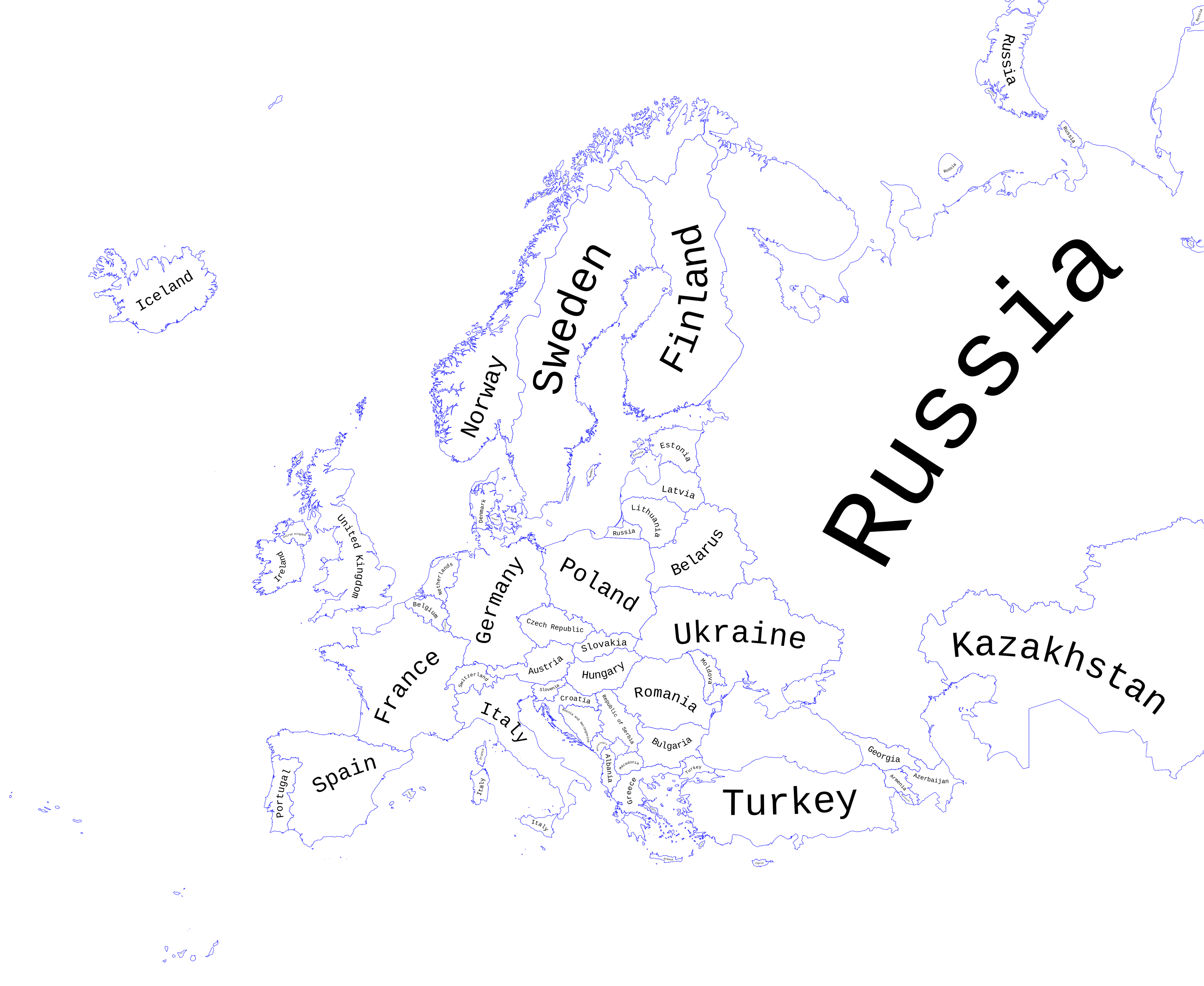}
        \caption{Political map of the European countries with Russia labelled statically as in figure \ref{fig:europeRuss} (top) and with a dynamic labelling (bottom).}
        \label{fig:europeCentral}
    \end{figure}

    A major challenge in interactive maps is that the places at the view port boundaries are intersected and incomplete.
    The corresponding static label may not be completely visible or not be visible at all (see Fig. \ref{fig:europeCentral} top).
    A dynamically computed label as displayed in Figure \ref{fig:europeCentral} (bottom) is much better -- see the Russia or Kazakhstan label.
    To allow for a pleasurable user experience the label computation needs to be very efficient.
    
    Considering interactive maps where no fixed zoom levels are rendered but the user is allowed to zoom continuously.
    It is difficult to determine in which setting an area can be labelled such that the label can be read well, especially for a large data sets.
    The automated placement of an area label of maximum size within the area allows to deduce this information.
    So it helps to automatically create digital, interactive maps with a high degree of detail.

    \subsection{Related Work}
    
    Eduard Imhof in 1975 (see \cite{imhof_1975}) systematically described what a good labelling of a map is like.
    The main goal is to provide a good readability, i.e. that the user can easily identify the feature corresponding to a label.
    Furthermore the labels need to be non-overlapping.
    The label should also reflect the importance and classification of its feature.
    And obviously the map should be labelled with a good density.
    Concerning area features he recommends to label it either with a horizontally aligned label or a curved label which is conform to the area shape.
    In both cases the label should be completely contained within the area and leave a free space of one to one-and-a-half of a character to the area boundary.
    In case of a curved label it should be based on a circular arc with a circular angle smaller then 60 degrees.

    In the most recent publication in this field, M. Barrault in 2001 proposed a set of criteria to measure the quality of an area label (\cite{barrault_2001}).
    When displaying an area label there are several degrees of freedom, i.e. differing inter-character, -word and -line spacings.
    These can be adopted to better fit the label into its corresponding area.
    In the same publication he is introducing an algorithm to compute area labels based on his proposed quality measures.
    This algorithms is based on the skeleton of the area which describes the shape and topology of the area.
    The concrete label position is constructed by approximating paths in the skeleton by circular arcs.
    These arcs are used to place a label and evaluate the positioning.
    The best of these candidates is used to label the area.
    Barrault in his work also describes some shortcomings of his approach.
    The most important one is related to the enumeration of the paths which are considered to be the most promising candidates to be further evaluated.

    In her bachelor thesis N. Mendel reimplemented and evaluated his results on real world area data of the German state \cite{mendel_2018}.
    She could reproduce the shortcomings and showed that best candidates are often not considered at all.
    She also pointed out that the quality measure formula in Barrault's work contains an error leading to unwanted results.
    Details and a fix are given in her work.

    \subsection*{Our Contribution}

    We aim to remedy the shortcomings of Barrault's algorithm.
    Our goal is to optimize the time required for computing a labelling to enable interactive labelling of maps.
    We consider a simplified label model with fixed font rendering parameters, i.e. to use fixed inter-character, -word and -line spacings.
    So the problem is reduced to placing a circular box (i.e. bent along a circular arc) with a given aspect ratio.
    This ratio $A$ describes the ratio of the height $H$ to the length $L$ of the bounding box of a label, i.e. $H = A \cdot L$.
    Our solution allows to compute labellings of $~120$ countries in less than 2 seconds.
    This implies near real-time for computing a labelling, which means the labelling does require significantly more time than the request of new map data over a (mobile) network connection.

    We use the same algorithmic pipeline as Barrault, but improve several steps to overcome the shortcomings (see section \ref{subsec:barrault}).
    The medial axis of the polygon is approximated based on the voronoi graph of the polygon vertices (section \ref{subsec:med_ax}).
    Within this skeleton we optimize the search for candidate paths to get a more diverse set of candidates and related circular arcs (section \ref{subsec:path_finding}).
    For the so computed arcs we provide an efficient method to find an optimal position of the label box along the arc (section \ref{subsec:lbl_placement}).
    We provide implementation details and experimental results (section \ref{sec:impl_exp}).

\section{Preliminaries}

    \subsection{Medial Axis}

        The \textbf{medial axis} of a planar shape was first defined by  \cite{blum_1967} and \cite{lee_1982} amongst others.
        Given a simple polygon $P$ representing the shape.
        The medial axis is defined as the locus of points $p$ internal to $P$ such that at least two points on the polygon's boundary are equidistant and closest to $p$. 
        This definition can be extended to polygons with holes in a straightforward manner.
        Each point on the medial axis can be assign a radius, describing the distance to the boundary (\cite{dey_2004}).

        \begin{figure}[ht]
            \centering
            \includegraphics[width=0.33\textwidth]{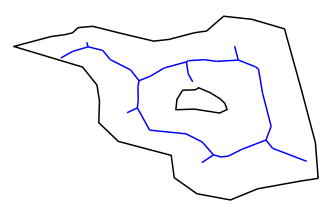}
            \caption{Approximated skeleton of a polygon with a hole.}
            \label{fig:skel}
        \end{figure}

        As \cite{schmitt_1989} and \cite{brandt_1994} point out the medial axis, also called skeleton, of a polygon can be approximated using voronoi diagrams.
        It is a special subset of voronoi edges namely those who are completely contained in the polygon (\cite{mcallister_2000}).
        See Figure \ref{fig:skel} for an example.
        For each of these we can approximate the minimum distance to the polygon boundary, what we will call its clearing (for details see section \ref{subsec:med_ax}).
        We are going to use this clearing to find paths through the skeleton-graph which offer a good amount of space to fit the label.

        \begin{figure}[ht]
            \centering
            \begin{tikzpicture}[scale=.4]
                \draw[rounded corners=3pt] (-1,0) -- (-1,.8) -- (0,2.5) -- (2,3.3) -- (5,3.5)
                    -- (6,3.3) -- (7,3.5) -- (8.5,3.3) -- (8.7,3.7) -- (8.2,4) -- (7.7,4.1)
                    -- (6,5) -- (5.5,5.5) -- (5.5,5.8) -- (5.7,6) -- (6.5,6.3) -- (8,6.2)
                    -- (9.7,5.8) -- (11,5.3) -- (12,4.5) -- (12.3,4) -- (12.5,3.5)
                    -- (12.2,1) -- (11.5,0) -- (10,-1) -- (9,-1.2) -- (8,-1.3)
                    -- (6,-1) -- (5,-1.1) -- (4,-1.2) -- (3,-1.2) --(0.5,-0.9)
                    -- (-0.8,-0.5) -- (-1,0);

                \begin{scope}[shift={(7,-6)}]
                    \draw[dashed, thin] (-8,0) arc (180:45:8);
                    \node[below] at (0,0) {$(x,y)$};
                    \draw[<->] (0, 0) --node[below]{$R$} (150:8);
                    
                    \draw[dotted] (0,0) -- (53:8);
                    \draw[dotted] (0,0) -- (141:8);
                    \draw[color=orange,thick] (141:8) arc (141:53:8);

                    \draw[dashed] (0,0) -- (65:8.5);
                    \draw[dashed] (0,0) -- ++(130:8.5);
                    \draw[ultra thick,color=green,opacity=1] (130:8) arc (130:65:8);

                    \draw[dashed,thick] (0,0) -- (120:14);
                    \draw[->] (53:2.25) arc (53:120:2.25) node[pos=.5,above] {$s$};
                    \draw[<->] (118:8) --node[right]{$l_d(s)$} (118:5.5);
                    \draw[<->] (121.5:8) --node[left]{$l_u(s)$} (121.5:10.6);
                \end{scope}
            \end{tikzpicture}
            \caption{An area label is bent along a support line (orange).
                The actual placement is the baseline (green).
                The quality of a placement can be measured by integrating the upper and lower distance to the area boundary along the baseline.}
            \label{fig:labelParams}
        \end{figure}
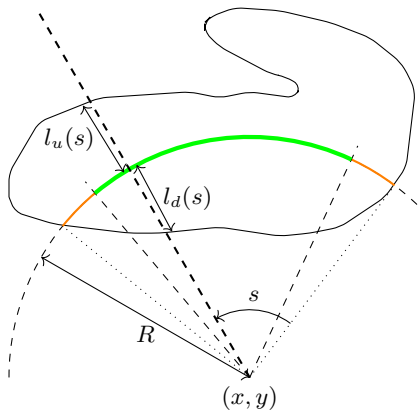

    \subsection{Quality Measures for Area Labels}\label{sec:label_quality}

        To measure the quality of an area label Barrault is evaluating six criteria.
        In the following the term \textit{longitudinal} is used to describe the left-right dimension and respectively \textit{latitudinal} for the top-bottom dimension.
        \begin{description}
            \item[Longitudinal extent:]
                The extent along the circular arc should be maximized.
            \item[Longitudinal centre:]
                The label should be centered in the polygon in the longitudinal dimension \dots
            \item[Latitudinal centre:]
                \dots as well as in the latitudinal dimension.
            \item[Conformity:]
                The base arc of the label should be conform to the shape of the labelled area. 
            \item[Orientation:]
                The more horizontally the label, the better.
            \item[Curvature:]
                A label based on an arc with larger radius is preferred.
        \end{description}

        For the following keep in mind Figure \ref{fig:labelParams} for an illustration.
        The center point and radius of a circular arc are defining the support line of a label (orange in Fig \ref{fig:labelParams}) - the line along which the label is bent.
        The possible label positions are bounded by the points where the arc is intersecting the polygon.
        A label position is determined by two angles describing start and endpoint of the label along the support line.
        This concrete position is called the \textit{baseline} of a label (green in Fig. \ref{fig:labelParams}).
        A valid baseline length is obliviously determined by the label length and the inter-letter and -word spacing.
        These are variable and in a general case can freely be chosen from given ranges.

        The quality of a label candidate is evaluated by its perceived coverage of the polygon.
        It is computed via an integral along the baseline, summing the minimum of the space above and below the label (named $l_u(s)$ and $l_d(s)$ in Fig \ref{fig:labelParams}).
        An addition a cost is induced if the endpoints of the label are too close to the border of the area.
        This is to prevent labellings from looking crammed.
        
        Barrault is considering the problem of placing a label of fixed font size within the area.
        To fit the label to the area shape he uses variable inter-letter and -word spacings.
        The underlying scenario is to label the area on a map with a fix map scale.
        This setting determines the font size of the label.

        When considering dynamic maps the focus point shifts slightly.
        Instead of finding a label position for a label of fixed font size, one aims for finding a label of maximum font size.
        Having computed such a label position allows to determine the map scales at which the label can be visualized appropriately.

        So our focus shifts to the following question:
        \begin{quotation}
            Considering a polygonal area, what is the label of maximum which can be place in the area such that the label fulfilles the criteria of a good labelling as defined above
        \end{quotation}

        To find this placement we suppose the inter-letter and -word spacings are fixed.
        With these values given, we can compute the bounding box of the label for a given font size.
        To meet the requirement of E. Imhof: to "leave a space at least one-and-one-half the size of the letters on either end of the word" \cite[p. 136]{imhof_1975}, we adapt the label accordingly.
        This gives us the length $L$ and height $H$ of the bounding box of the label.
        From these we can compute the ratio $A = \frac{H}{L}$, i.e. the ratio of the height to length of a label of arbitrary font size.
        Our goal is to place a box of this aspect ratio within the polygon, such that:
        \begin{itemize}
            \item The box is bent along a circular arc (preferably with a large radius).
            \item The longitudinal is maximized.
            \item The label is centered in longitudinal as well as latitudinal dimension.
            \item It should be conform to the shape of the area
        \end{itemize} 
        
        In the following section we will introduce our algorithm to efficiently compute such an optimal box position and size.
        The algorithm is based on the idea provided by Barrault but avoids some of the drawbacks of his approach and hence achieves much better performance.
        So we are able to compute area labels very efficiently in near-real time.

        With this algorithm at hand we can think of further methods to the labelling of areas in interactive settings.
        Consider the scenario where an area is only partially visible, i.e. a larger part of the area is not contained within the current map view.
        Using fixed area labels causes problems if the label is mainly placed in the invisible part of the area.
        Being able to quickly compute area labels makes allows for labelling the visible area only.
        Alternatively the labeled area might be cropped to reach into the invisible area.
        This might indicate to the user that the area extends into the invisible part, while the main part of the label is readable.
        The whole area and label can then be interactively explored by panning the map accordingly. 
        
        Beyond that the computed label allows to determine visualization settings where the area label can be visualized properly.
        This allows for improving the automated generation of high quality interactive maps.

\section{Medial-Axis-based Curved Area Label Placement}

    \subsection{High-level Idea}

    As we want to place our labels along a circular arc, we first have to find some arcs, which are a reasonable fit to the polygon.
    To find such an arc we use an approximation of the medial-axis.
    A long path through this graph should be an appropriate representative of the area's shape.
    Because we want our labels to be placed along circular arcs we fit a circle through the vertices of the path.
    Multiple candidates are enumerated and evaluated to find the best placement along each arc.
    Of the so obtained labellings the largest is reported.

    \subsection{Barrault's Incarnation}
    \label{subsec:barrault}

    Barrault's algorithm follows the steps described above.
    To decrease the complexity of the input polygon morphologic erosion is applied.
    For the eroded polygon a delaunay triangulation is computed.
    For each of the delaunay triangles a convex-combination of its corners defines a ``center point'' of the triangle.
    Those ``center-points'' of adjacent triangles are connected and thus form the edges of the medial-axis approximation.

    After approximating the medial-axis in this manner, the $50$ longest shortest paths are considered as candidates.
    A circle is fitted through each of them and investigated further as a possible label support line.

    To find an optimal label placement for each support line, all possible placements (discretized) are considered.
    That is every possible combination of starting and ending angle.
    For each placement the label is evaluated.
    The placement with the highest score is then returned as the optimal label.

    A major drawback of Barrault's approach is his choice of the 50 paths he is evaluating.
    These paths are mostly very similar and so are the fitted circular arcs as Mendel shows in \cite{mendel_2018}.
    As a result many promising alternatives are not considered at all.
    Additionally the evaluation of the possible label placements contain several integral computations. 
    Each requiring much computation power.
    Overall the computation takes a long time and in many cases does not even leads to optimal results.

    \subsection{RALF -- Real-time Area Label Fitting}

    We go beyond Barrault's algorithm in several points.
    Firstly we use a medial axis approximation based on the voronoi graph.
    This allows for each edge in the medial axis to approximate the minimum distance to the boundary polygon.
    We call this distance the clearance of an edge.

    The clearance value is used to find paths in the skeleton which are promising to fit a label through.
    This discards paths that are to close to the border of the area, which would restrict the label size.

    We also improve path selection by computing a more diverse set of paths.

    A third improvement is related to computing an optimal position of the label along a candidate arc.
    Here we are proposing a new scheme to compute an optimal placement along the arc.

    \subsubsection{Medial Axis Approximation}
    \label{subsec:med_ax}

\begin{figure}[ht]
        \centering{
            \includegraphics[width=0.48\textwidth]{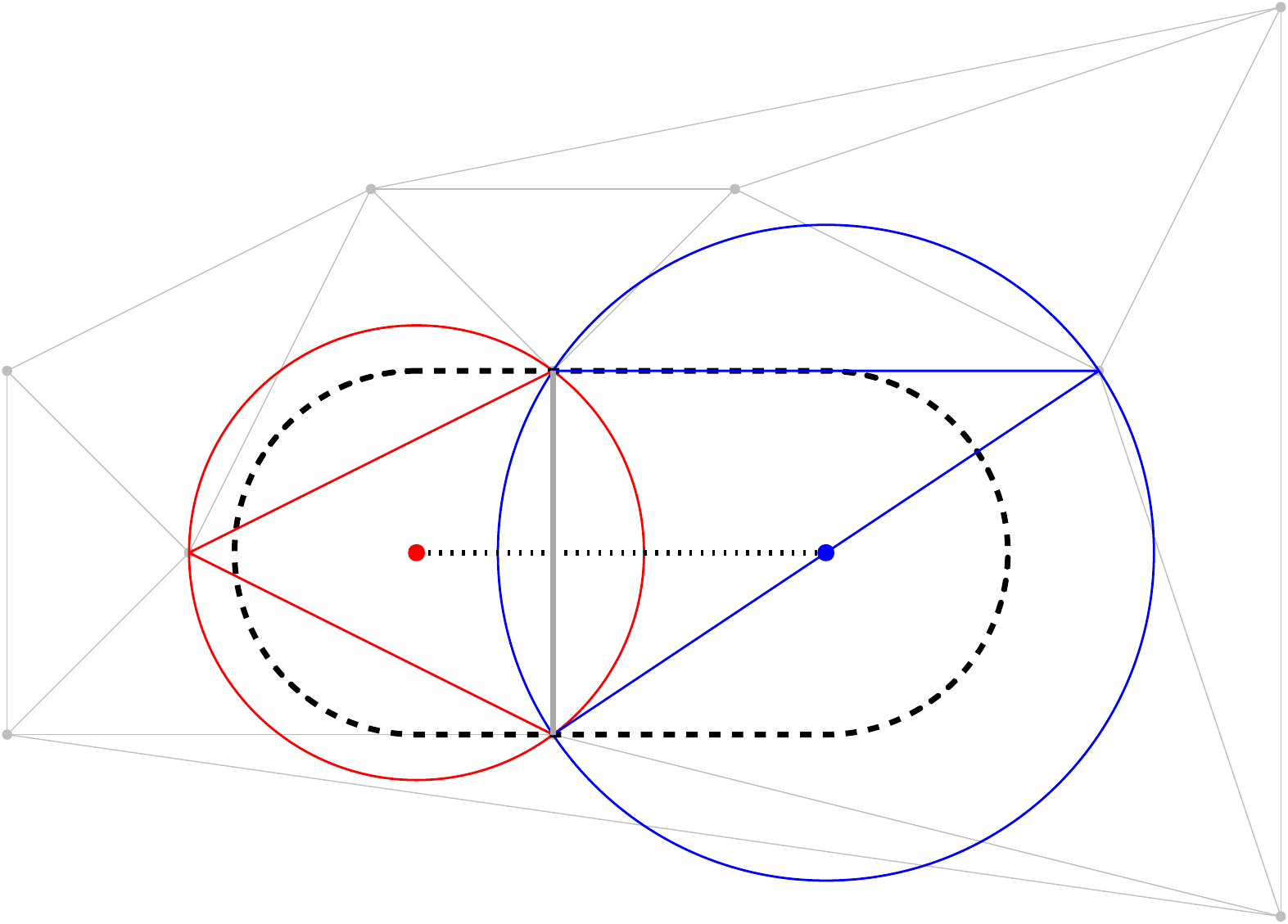}
            \hfill
            \includegraphics[width=0.33\textwidth]{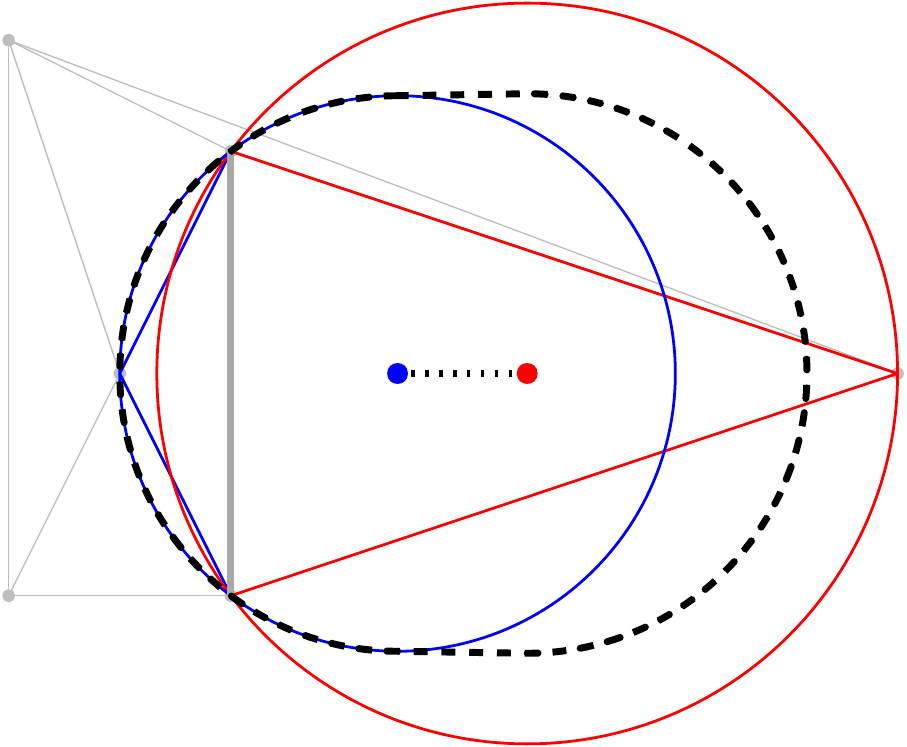}
        }
        \caption{The clearance of a voronoi edge if the centers are on the same side (top) or on different sides (bottom) of the delaunay edge.}
        \label{fig:clearance}
    \end{figure}

    To get an approximation of the medial-axis where we are able to bound the distance to the boundary polygon we proceed as follows:
    We compute the delaunay triangulation of the boundary polygon.
    For each delaunay triangle the voronoi center is defined as the center of the circumcircle of the delaunay triangle.
    We connect the voronoi centers of adjacent delaunay triangles if this so called voronoi edge is completely contained within the polygon.
    For these edges we can approximate the clearance, i.e. the minimum distance to the boundary polygon.
    We need to distinguish two cases:
    If \textit{the voronoi centers are on different sides of the delaunay edge} the minimum clearance is half the length of the radical line of the two circumcircles (i.e. the delaunay-edge itself).
    In the second case \textit{both voronoi centers are located on the same side of the delaunay edge}.
    Then the clearance is the minimum of the radii of the two corresponding circumcircles.
    See Figure \ref{fig:clearance} for an example.

    The rational behind this is as follows:
    All points closer to the segment than the computed distance are also contained in at least one of the voronoi-balls.
    Those balls are empty of other points by definition of a valid delaunay triangulation.
    Therefore there are no points within the cleared segment.
    The clearance of the segment might still intersect the boundary of the polygon.
    But this can be remedied if the boundary is sampled sufficiently dense.
    Furthermore we only use these clearance-values as guidance but do not rely on them for correctness (i.e. keeping the labelling within the polygon).

    The so constructed skeleton graph has an associated clearance value for each of its edges.

    \subsubsection{The Algorithm}

    Having constructed the skeleton graph and the clearance values one is after finding promising paths in the skeleton.
    This paths should allow to place a label of maximum size.

    \subsubsection{Finding Candidate Paths}
    \label{subsec:path_finding}

    We aim to find a set of $k$ diverse candidate paths which we further investigate to place a good label.
    Our strategy is based on the following observation:
    If we place a label along a given path, the minimum clearance of the path-edges hints at the maximum possible height of the label along this path.
    We therefore are looking for paths with a high minimal clearance, whose length allows to fully utilize the vertical space promised by this clearance.
    That is the length of the path should be no less than $l_{min} = \frac{2*clearance}{aspect}$.

    The idea is to start with a large clearance value (e.g. the maximum clearance value) and remove all edges of the skeleton which have smaller clearance value.
    In this subgraph we search for shortest paths such that their length is larger than the appropriate minimum length.
    If we can't find enough paths, we reduce the clearance and search for the remaining paths in the subgraph filtered with the new clearance.
    In our case we start with the maximum clearance in the graph and reduce it by $\sqrt{2}$, i.e. we half the area of the label box we search for.

    In detail we proceed as follows:
    In each component of the pruned skeleton we start with an arbitrary node and search for the node which is furthest away.
    This is done with one dijkstra call by tracking the root node of every shortest-path-tree.
    The so found nodes form our set of start nodes.
    We now search for the node which is furthest from our set of start nodes - also requiring only one dijkstra call with all the nodes as sources.
    The so found pair of nodes approximates the longest shortest path in the pruned skeleton (this method is exact for trees but not for arbitrary graphs).
    If the path length is larger than $l_{min}$ we report the path and add its vertices to the set of start-nodes.
    If we did not yet find $k$ paths, we repeat the search with the new set of start-nodes.
    If the found path is of shorter length, we decrease our clearance by $\sqrt{2}$, refilter the graph and proceed as described.
    We repeat this until we found the $k$ paths.

    Through each of the candidate paths a circle is fitted.
    Let $p_1, \dots p_n$ be the points of the path.
    We compute a center $c$ and a radius $r$ such that the term $\sum_{i=1}^{n}\left( \left| p_i - c\right| - r\right)^2$ is minimized.

    \subsubsection{Label Placement}
    \label{subsec:lbl_placement}

    \begin{figure}[ht]
        \centering
        \includegraphics[width=0.6\textwidth]{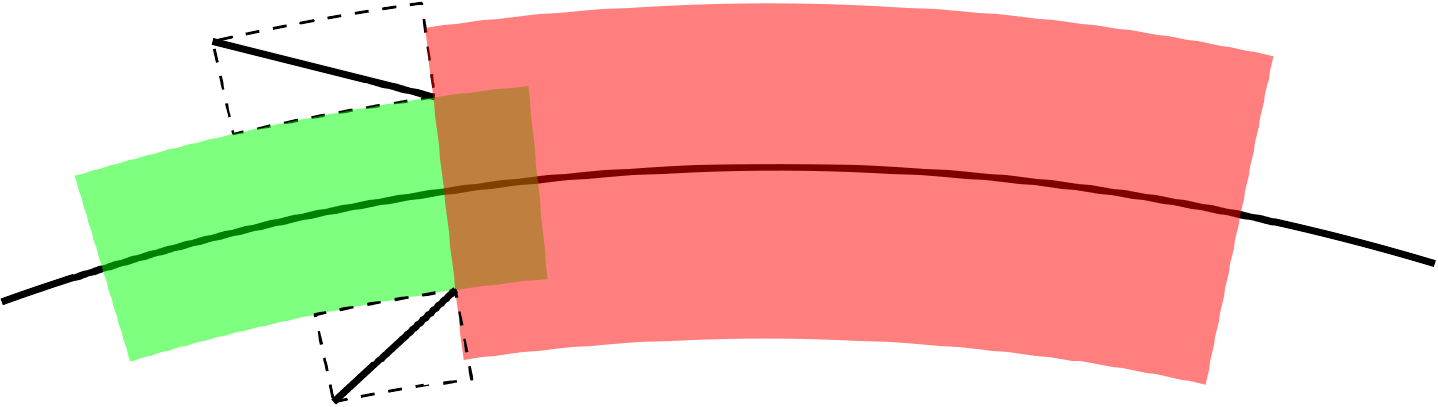}
        \caption{
            The segments of the polygon restrict the label size.
            If the label is placed below or above a segment, the segment constrains the possible size (green label).
            We have to move the label considerably to the side so it can grow (right label).
        }
        \label{fig:wedge_3lin}
    \end{figure}

    Given a circle, a polygon and a text label we aim to find the position along the arc such that the size of the label can be maximized.
    We can compute this optimal placement in time $n\log n$, where $n$ denotes the size of the polygon.

    Let us first consider how a single polygon-segment constrains the label placement.
    We employ the following simplification: A circular bounding-box is constructed around each polygon-segment.
    
    There are two cases:
    First if the label size is restricted by the segment in its height, then we can move it along the arc without getting any benefit.
    In the second case the size of the label is restricted by the segment in its length.
    In this case the size of the label increases if we move the center of the label away from the segment.
    The more we shift it away from the center, the more we can increase the size of the label.
    If we consider the possible size of the label as a function of the angle where the label is placed on the circle, we get a piecewise function with $3$ parts:
    When the label gets closer to the segment the possible width decreases until it can fit below/above the segment.
    It then stays constant, while passing above/below the segment and finally increases.
    Let's call those functions ``wedges''.
    For a given angle, they tell us how large a label can be if it is placed at this angle on the circle. 
    This is illustrated in Figure~\ref{fig:wedge_3lin}.

    We now construct all those ``wedges'' from the segments and find the highest point, which is below each of the wedges.
    This point describes the position where the label width is maximal. 
    Because the label-aspect is fixed this means that the overall label size is maximized.
    So this yields the optimal label placement.

    \begin{figure}[ht]
        \centering
        \includegraphics[width=0.32\textwidth]{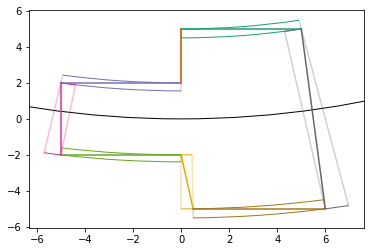}
        \hfill
        \includegraphics[width=0.32\textwidth]{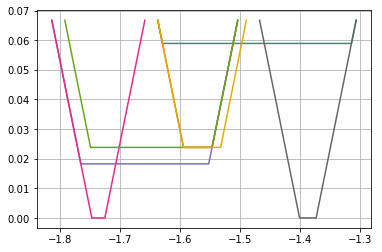}
        \hfill
        \includegraphics[width=0.32\textwidth]{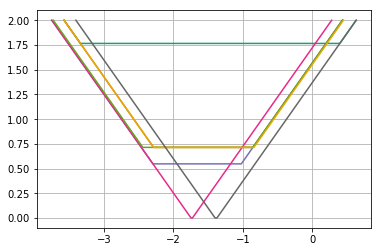}
        
        \caption{Polygon segments with their circular bounding boxes (top) and the corresponding bounds in the circular diagram for a very tall label (middle) and a very long label (bottom).
            The tall label is constrained by the cyan segment.
            We can actually move it a little to left or to the right.
            The long label is constrained by the pink and the gray segments.
        }
        \label{fig:wedge}
    \end{figure}

    To find this point, we first consider the complete circle from $0$ up to $2\cdot\pi$.
    We then consider each wedge, from lowest to highest, and restrict the possible placements.
    When there are no more valid placements left we return the highest point seen.
    A simple example instance is depicted in Figure \ref{fig:wedge}.

    The active wedges can be organized in a segment tree.
    For any height, the set of wedges looks like a set of segments.
    When going up, these segments grow.
    When two wedges intersect, we can merge the associated segments.

    We can enumerate the wedges with a heap.
    Because we can stop the computation, when there are no more valid placements left, we only consider a small amount of segments.
    Considering the example in Figure \ref{fig:wedge} with the long label, we would only inspect the pink and gray wedges before returning the optimal placement.

    \subsubsection*{Wedge Computation}

\begin{figure}[ht]
        \centering
        \includegraphics[width=0.33\textwidth]{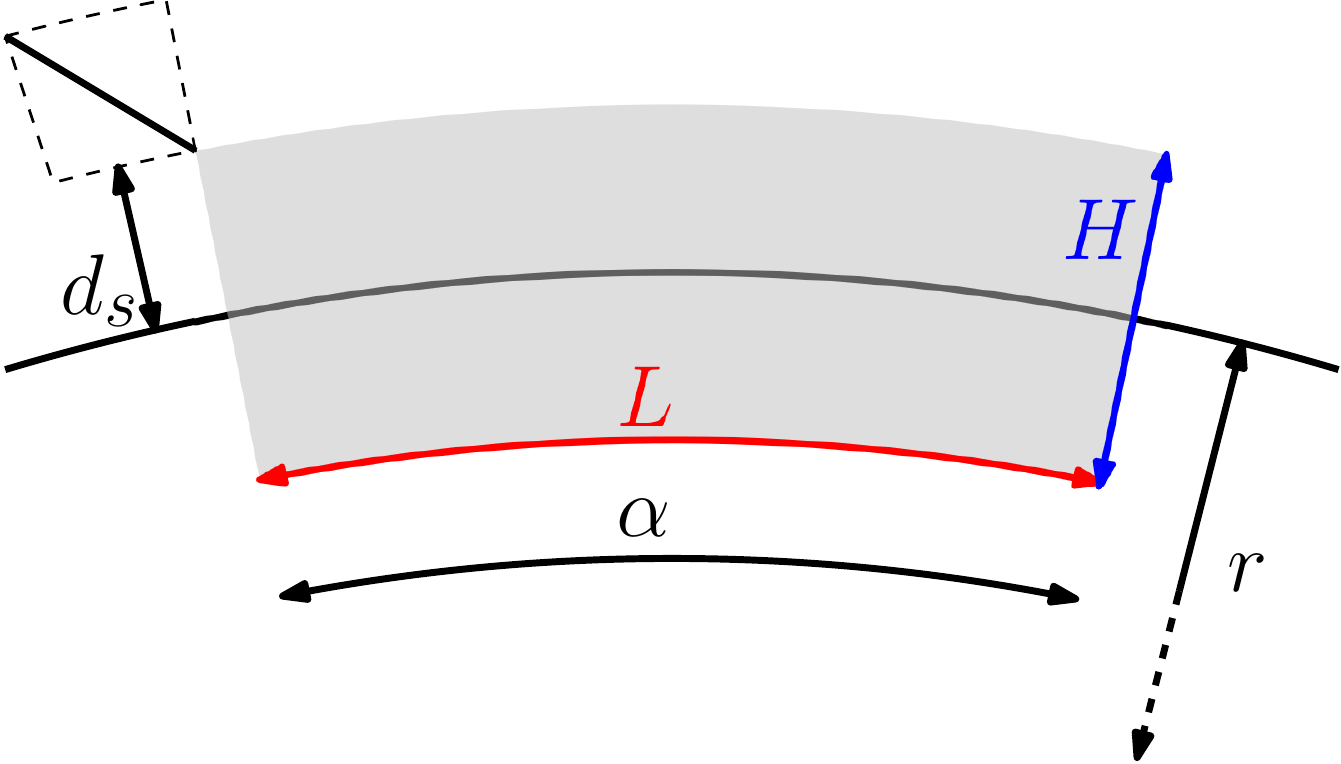}
        \caption{The label can start to grow, when its corner touches the corner of the segment's bounding box.}
        \label{fig:wedge_comp}
    \end{figure}

    In this chapter we will go through the math needed to actually compute the wedges.
    For any mathematical symbols, please consider Figure \ref{fig:wedge_comp} as a reference.
    Furthermore $A$ denotes the aspect of the label.

    First let's derive the relationship between the height of the label ($H$), its width ($L$) and the spanned angle ($\alpha$) for a given circle with radius $r$.

    By definition the following holds: 
    \begin{equation}\label{eq:wedge1}
        H = L\cdot A
    \end{equation}
    Furthermore we can easily derive:
    \begin{equation}\label{eq:wedge2}
        L = (r - H/2)\cdot\alpha
    \end{equation}

    For a given segment $s$ let $d_s$ denote the minimal distance of the segment to the circle.
    If the height $H$ of our label is less than $d_s$ the segment does not interfere with the label placement.
    If the height $H$ of the label is greater than $d_s$, we can compute the spanned angular range $\alpha$ by plugging \eqref{eq:wedge2} into \eqref{eq:wedge1} and solving for $\alpha$.
    The center of the label needs to be at least $\alpha/2$ from the segment.

    With the special case of $H = d_s$ we can compute exactly the placement of the label for which the wedge transitions from one linear function to the next.
    Coming from the left the label shrinks, until it can fit below the segment.
    It then slides along without changing size.
    Finally its size can increase once again if its far enough to the right.

    The following statements are equivalent:
    \begin{itemize}
        \item The height of the label is maximized.
        \item The length of the label is maximized.
        \item The area of the label is maximized.
        \item The angular extent of the label is maximized.
    \end{itemize}
    It's easiest to describe the wedges in terms of maximum angular extent.

    Let $\alpha_{d_s}$ be the alpha value, such that $H$ equals $d_s$.
    Also let $\beta_1$ and $\beta_2$ be the angles between the circle center and the segment's endpoints.
    Finally let $\alpha_l$ denote the angle on which the label center is placed.

    If $\alpha_l > \beta_2 + \alpha_{d_s}/2$ the maximum possible label extent is $\alpha_{d_s} + 2\cdot\left( \alpha_l - \left( \beta_2 + \alpha_{d_s}/2\right) \right)$

    If $\alpha_l < \beta_1 - \alpha_{d_s}/2$ the maximum possible label extent is $\alpha_{d_s} + 2\cdot \left( \left( \beta_1 - \alpha_{d_s}/2 \right) - \alpha_l \right)$

    If $\alpha_l$ falls within those bound the value is exactly $\alpha_{d_s}$.

    This yields 3 piecewise linear functions for the wedges.

\section{Implementation and Experimental Results}
\label{sec:impl_exp}

    \subsection{Implementation}
    We implemented our algorithm in \texttt{C++}.
    For the geometric operations we relied on the CGAL Library~\cite{CGAL}.
    Graph searches were done with the help of the Boost Graph Library~\cite{BoostLibrary}.
    The code was compiled with gcc 8.3.
    The Experiments were run on a standard desktop computer with a Intel Xeon E3-1225v3 CPU with 3.20GHz.

    \subsection{Benchmarks}

\begin{figure}[ht]
        \centering
        \includegraphics[width=0.47\textwidth]{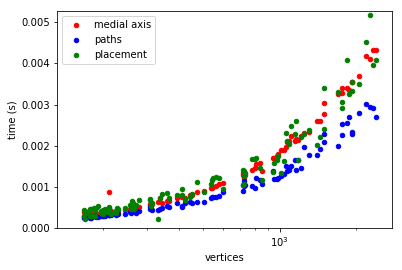}

        \includegraphics[width=0.47\textwidth]{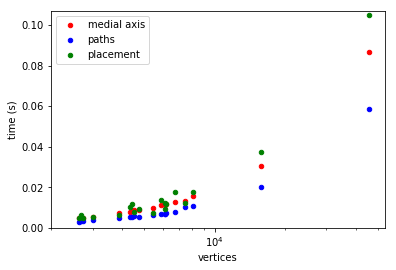}
        \caption{Running time of the main operations on different sized data-sets.
        The diagram is split in two parts along the x-axis to improve readability.
        Also mind the log-scale along the x-axis.}
        \label{fig:eval}
    \end{figure}

    We evaluated our code on a data set of all 50 European countries including Russia.
    The data-set was obtained from \cite{NaturalEarth}.
    For some of the countries multiple polygons were provided to accommodate small islands belonging to those countries.
    The complete polygon set consisted of $220,000$ nodes.
    While the smallest areas where not labelled, the set of labelled areas was of size 120.
    We computed all labels in just under 2 seconds.
    The running times for the main operations are depicted in Figure \ref{fig:eval}.

    Running times for the different operations are very close.
    When inspecting the work-per-node in table~\ref{tab:eval}, one can see that the running-times are tightly coupled to the number of nodes. 
    The spread between the running times (work-per-node) is at most 4, which fits our data set as the largest polygon is about 4 orders of magnitude larger than the smallest.

    \begin{table}
        \centering
        \begin{tabular}{ r | c c c}
        &             & path-       &           \\
        & medial axis & enumeration & placement \\
        \hline
        mean ($\mu$s) & 1.86 & 1.39 & 1.96 \\
        std ($\mu$s) & 0.22 & 0.14 & 0.35 \\
        spread & 2.57 & 1.66 & 4.17 \\
        (max/min)
        \end{tabular}
        \caption{Running times averaged over the number of nodes.}
        \label{tab:eval}
    \end{table}

\section{Conclusions}

    We presented an algorithm to automatically label areas with curved labels.
    The input areas are given as a boundary polygon which might contain holes.
    Our proposed algorithm allows to compute labellings for large data sets in just a few seconds.
    So a labelling for the European countries which consists of $124$ separately labelled polygons containing around $220,000$ nodes takes $2$ seconds.

    Our algorithm is a refinement of the algorithm Barrault proposed in 2001.
    Instead of placing an label with fixed font size, like Barrault does, we aim for finding a label position that allows to display the label as large as possible.
    We improve over Barrault's algorithm by only considering paths, which offer a reasonable amount of free space around them.
    We then improve the path selection to compute a more diverse set of candidate paths, while requiring fewer candidates.
    Finally we compute the optimal placement of the label along the arc with a simple but efficient algorithm instead of brute-forcing all possible placements.

    \subsection{Outlook}

    Our algorithm performs quite well on real world data sets and produces very satisfactory results.
    Nevertheless the algorithm could be further improved by taking the variable spacings into account.
    This would affect the search of the candidate-paths in the skeleton.
    As the minimum length of the considered paths would depend on the maximum possible aspect.
    In particular it would concern the placement of the label along the candidate circle.
    If the aspect of the label is variable, so is the ascend of the wedges.

    We considered the input data in full resolution. 
    But we could also reduce the number of nodes as described in \cite{mendelarea18}.
    This should speed-up the algorithm considerably, by reducing the number of nodes.
    If the error-threshold is chosen small enough, the quality should not suffer noticeably.
    On the other hand, one has to make sure, that the boundary keeps sampled evenly enough.

\bibliographystyle{abbrv}
\bibliography{references.bib}

\begin{thebibliography}{10}

\bibitem{NaturalEarth}
{Natural Earth}.
\newblock \url{https://www.naturalearthdata.com}, 2019.

\bibitem{barrault_2001}
M.~Barrault.
\newblock A methodology for placement and evaluation of area map labels.
\newblock {\em Computers, Environment and Urban Systems}, 25(1):33--52, 2001.

\bibitem{blum_1967}
H.~Blum.
\newblock A transformation for extracting new descriptions of shape.
\newblock {\em Models for the perception of speech and visual form}, pages
  362--380, 1967.

\bibitem{BoostLibrary}
Boost.
\newblock {Boost C++ Libraries}.
\newblock \url{http://www.boost.org/}, 2019.

\bibitem{brandt_1994}
J.~W. Brandt.
\newblock Convergence and continuity criteria for discrete approximations of
  the continuous planar skeleton.
\newblock {\em {CVGIP}: Image Understanding}, 59(1):116--124, 1994.

\bibitem{dey_2004}
T.~K. Dey and W.~Zhao.
\newblock Approximating the medial axis from the voronoi diagram with a
  {ConvergenceGuarantee}.
\newblock {\em Algorithmica}, 38(1):179--200, 2004.

\bibitem{imhof_1975}
E.~Imhof.
\newblock Positioning names on maps.
\newblock {\em The American Cartographer}, 2(2):128--144, 1975.

\bibitem{lee_1982}
D.~T. Lee.
\newblock Medial axis transformation of a planar shape.
\newblock {\em {IEEE} Transactions on Pattern Analysis and Machine
  Intelligence}, {PAMI}-4(4):363--369, 1982.

\bibitem{mcallister_2000}
M.~{McAllister} and J.~Snoeyink.
\newblock Medial axis generalization of river networks.
\newblock {\em Cartography and Geographic Information Science}, 27(2):129--138,
  2000.

\bibitem{mendel_2018}
N.~Mendel.
\newblock Dynamische beschriftung von gebietshierarchien - entwicklung und
  implementierung der beschriftung von hierarchischen gebietsunterteilungen.
\newblock Bachelorthesis, University of Stuttgart, Stuttgart, 2018.

\bibitem{mendelarea18}
T.~Mendel.
\newblock Area-preserving subdivision simplification with topology constraints:
  Exactly and in practice.
\newblock In {\em Proceedings of the Twentieth Workshop on Algorithm
  Engineering and Experiments, {ALENEX} 2018, New Orleans, LA, USA, January
  7-8, 2018.}, pages 117--128, 2018.

\bibitem{schmitt_1989}
M.~Schmitt.
\newblock Some examples of algorithms analysis in computational geometry by
  means of mathematical morphological techniques.
\newblock In J.~D. Boissonnat and J.~P. Laumond, editors, {\em Geometry and
  Robotics}, Lecture Notes in Computer Science, pages 225--246. Springer Berlin
  Heidelberg, 1989.

\bibitem{CGAL}
{The CGAL Project}.
\newblock {\em {CGAL} User and Reference Manual}.
\newblock {CGAL Editorial Board}, {4.14} edition, 2019.

\end{thebibliography}
\end{document}